\documentclass[twocolumn,amsmath,amssymb,aps]{revtex4}

\usepackage{graphicx}

\begin{document}

\title{Synchronization by dynamical relaying in electronic circuit arrays}

\author{Iacyel Gomes Da Silva$^1$, Javier M. Buld\'u$^2$, Claudio R. Mirasso$^1$
and Jordi Garc\'{\i}a-Ojalvo$^3$}

\affiliation{
$^1$Departament de F\'{\i}sica, Universitat de les Illes Balears,
E-07122 Palma de Mallorca, Spain\\
$^2$Nonlinear Dynamics and Chaos Group,
Departamento de F\'{\i}sica Aplicada y
Ciencias de la Naturaleza, Universidad Rey Juan Carlos, Tulip\'an s/n,
28933 M\'ostoles, Madrid, Spain.\\
$^3$Departament de F\'{\i}sica i Enginyeria Nuclear, Univ.
Polit\`ecnica de Catalunya, Colom 11, 08222 Terrassa, Spain.\\
}

\begin{abstract}
We experimentally study the synchronization of two
chaotic electronic circuits whose dynamics is relayed by a third
parameter-matched circuit, to which they are coupled bidirectionally in
a linear chain configuration. In a wide range of operating parameters,
this setup leads to synchronization between the outer circuits,
while the relaying element remains unsynchronized. The specifics of the
synchronization differ with the coupling level: for low couplings a
state of intermittent synchronization between the outer circuits coexists with
one of antiphase synchronization. Synchronization becomes in phase for
moderate couplings, and for strong coupling identical synchronization
is observed between the outer elements, which are themselves synchronized
in a generalized way with the relaying element. In the latter situation,
the middle element displays a triple scroll attractor that is not possible
to obtain when the chaotic oscillator is isolated.
\end{abstract}

\maketitle

\noindent
{\bf Synchronization of chaos between pairs of coupled dynamical elements has been
extensively studied in the last decade. With the recent surge of interest in
the collective behaviour of dynamical networks, it has become necessary to
determine the conditions in which synchronization persists in the presence of
multiple interactions between manifold oscillators. A first nontrivial extension
of the two-oscillator case is a linear array of three coupled elements. Here we
study experimentally such an architecture with nonlinear electronic elements
acting as nodes of the array. Our results show that a pervasive dynamical regime
of this system is one in which the outer elements of the chain are synchronized,
while the middle one is not, acting merely as a {\em dynamical relay} between 
the two other elements. Depending on the coupling strength, a rich variety
of synchronized regimes is observed.
}

\section{Introduction}

Synchronization between pairs of chaotic systems has been
studied profusely in recent years \cite{boccaletti02} in fields such as optics \cite{uchida05}, driven by potential technological applications, e.g. in secure
communications \cite{donati02,argyris05}. Nonlinear electronic circuits, in particular,
have boosted the study and understanding of chaos
synchronization, due to their simplicity and the fact that all
variables of the circuit are accessible and measurable.
Electronic circuits were used, for instance, in pioneering studies
on chaotic communications \cite{kocarev92,cuomo93}.

In this paper we use a particular nonlinear electronic circuit,
known as Chua's circuit or double-scroll oscillator
\cite{madan93}, which has become a paradigmatic example of chaotic
circuit. Chua's circuit exhibits a rich set of dynamical regimes,
ranging from stable to chaotic dynamics,
including periodic and excitable behaviors. Moreover, the circuit can be
modelled in a straightforward way by a set of three ordinary
differential equations. Chaos synchronization between pairs of these
circuits has been studied for both unidirectional \cite{parlitz00,zhu01} and
bidirectional \cite{canna02} coupling.

Recently, a large interest has appeared in the collective behavior of
networks of dynamical elements \cite{boccaletti06}. In this context, it is
necessary to determine how the network architecture determines the
synchronized behavior of coupled chaotic oscillators.
Synchronization in large arrays of electronic circuits has already been studied
\cite{lorenzo96,munuzuri99}. But even the simplest
extension of the standard two-element system, namely a linear array of
three coupled elements, exhibits nontrivial synchronization scenarios.
In this paper we investigate experimentally the
synchronization regimes of three Chua's circuits coupled
in a linear chain.
Coupling is introduced in a bidirectional way, allowing the
transfer of information in both directions. In this way, there is
not a clear leader in the dynamics neither a system that acts as a
simple follower. All circuits influence, in a certain way, the
dynamics of the others.

This kind of configuration of three interacting elements has been
investigated both theoretically \cite{winful90} and
experimentally \cite{terry99,fischer06} in coupled laser systems. 
Our experimental setup allows a systematic study of the
role of coupling in the collective behavior of the system. Our results
show that the coupling strength controls different synchronization regimes
between the circuits. In all the regimes observed, the central circuit acts as
a relay of the dynamics between the two outer ones. In particular, as will be
described in detail below, we observe:
a) For low coupling levels, intermittent and antiphase synchronization
between the outer elements coexist, depending on the initial conditions.
b) For moderate coupling, the outer circuits synchronize in phase, their
chaotic dynamics being relayed by
the central one, which seems to remain unsynchronized.
c) For large couplings, the outer lasers are synchronized to each other
identically, and to the central one in a generalized way. In that
regime, the central circuit exhibits a triple scroll attractor.


\section{The experimental setup}

\begin{figure}[htb]
\centering
\includegraphics[width=70mm,clip]{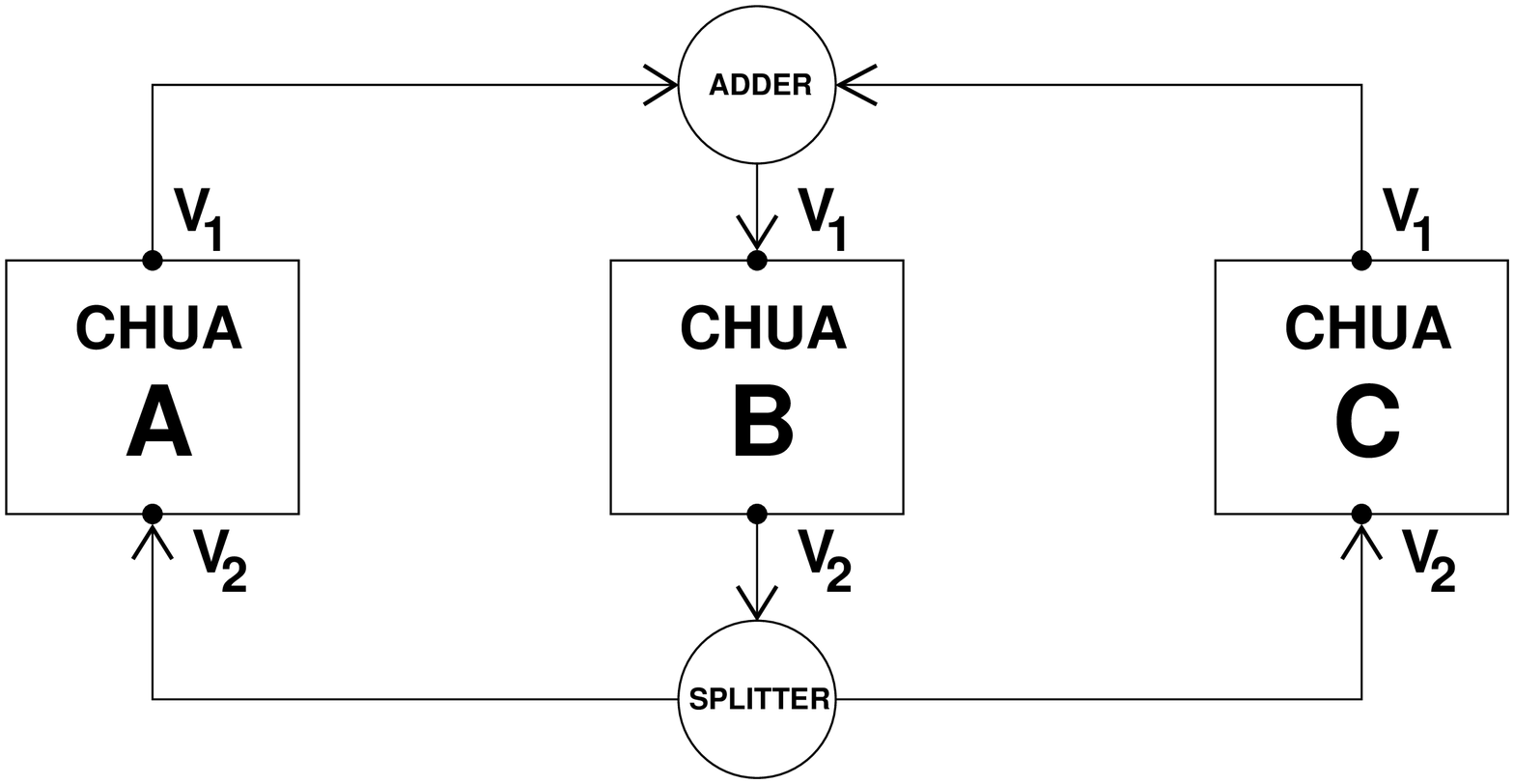}
\centering
\caption[2]{Schematic representation of the experimental setup.}
\label{fig:f01}
\end{figure}

We focus on the synchronization of three Chua's
circuits coupled in an open-chain configuration.
A detailed description of the circuit, with all its
components and connections and its corresponding rate equations,
is given in Appendix~\ref{app:circ}. As depicted schematically in Fig.~\ref{fig:f01},
a central circuit (B) is coupled to two outer ones (A and C).
Voltages $V_1$ or $V_2$ are sent to the nearest
neighbors via a voltage follower, in such a way that a bidirectional link is built 
upon two unidirectional lines. In other words, each circuit sends out
the state of one of its variables, but at the same time receives an
input through the other variable (see Appendix~\ref{app:coup} for details). 
In this way the circuits are bidirectionally coupled, with the central circuit
mediating between the external ones. It is worth noting that the central
unit is receiving (sending) two inputs (outputs) while the outer
units only receive (send) one. In this sense, one could reasonably
expect different dynamics between the central circuit and
the outer ones.

The coupling strength is
adjusted by a coupling resistance placed between the output of the
voltage follower and the input of the receiver circuit. We use a data acquisition card (DAQ) in order to
measure the value of $V_1$ and $V_2$ of each circuit. The DAQ card
has a sampling frequency of $50$~KS/s per channel and a maximum input voltage of $\pm10$~V.

\begin{figure}[htb]
\centering
\includegraphics[width=80mm,clip]{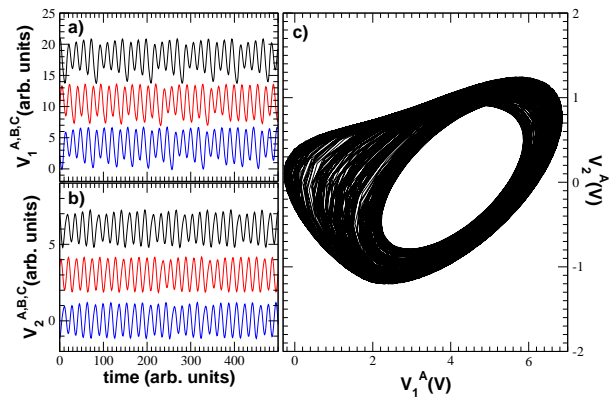}
\centering
\caption[2]{Temporal evolution of $V_1$ (a) and $V_2$ (b)
for the three circuits when
coupling strength is set to zero ($R_{\rm coup}\sim\infty$, see
Appendix~\ref{app:circ} for
details about the circuit design and Appendix~\ref{app:coup} for
coupling connections). From top to bottom: A (black), B (red)  and C (blue).
Voltages have been shifted vertically, in order to ease comparison between time series.
(c) Dynamics of circuit A in the phase space given by ($V_1$,$V_2$).
Circuits B and C have similar phase portraits (not shown here).
}
\label{fig:f02}
\end{figure}

We tune the internal parameters of the circuits so that they exhibit
chaotic dynamics in the absence of coupling. Figure~\ref{fig:f02} displays
the evolution of the output variables $V_1$ and $V_2$ of 
each isolated circuit [Figs.~\ref{fig:f02}(a) and (b)]
and the corresponding phase-space trajectories in the ($V_1$,$V_2$) plane
(Fig.~\ref{fig:f02}c). The dynamics of the isolated circuits corresponds to
a single-scroll chaotic attractor.

\section{The influence of coupling strength}

\begin{figure}[htb]
\centering
\includegraphics[width=60mm,clip]{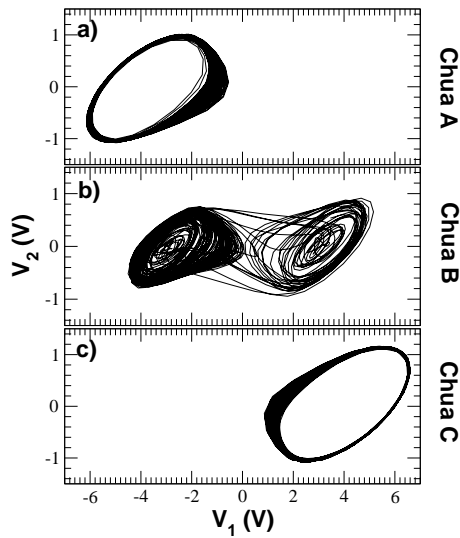}
\centering
\caption[2]{
Phase space dynamics of circuits A, B and C for low coupling, corresponding
to a coupling resistance $R_{\rm coup}= 20$~k$\Omega$.}
\label{fig:f03}
\end{figure}

Our aim is to study the influence of coupling strength in the
synchronization of the circuit array. Therefore, starting with the
three circuits uncoupled, we increase the coupling strength by
decreasing the value of the coupling resistances
$R_{\rm coup}^{A,B,C}$ placed at the input of each circuit. We
keep $R_{\rm coup}^A=R_{\rm coup}^B=R_{\rm coup}^C$ throughout the paper,
in order to guarantee the same amount of coupling between all
circuits. When $R_{\rm coup}=20$~k$\Omega$ we observe how the circuit
outputs begin to modify their dynamics. Specifically, two
different behaviors arise, depending on the initial
conditions of the array. Figure~\ref{fig:f03} plots the
trajectories in phase space for one of the two possible states. In this
case, the two outer circuits (A and C) oscillate in a
tight single-scroll attractor, although
around two different unstable fixed points. At the same time, the
central circuit (B) operates in a double-scroll
chaotic attractor. Changes in the dynamics induced by coupling were expected,
since they have been recently reported in two
bidirectionally coupled Chua's circuits \cite{gomes06}.

\begin{figure}[htb]
\includegraphics[width=80mm,clip]{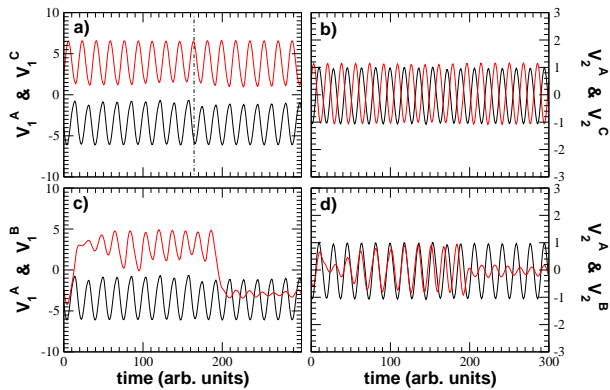}
\caption[2]{
Temporal evolution of $V_1$ (a) and $V_2$ (b) of the outer circuits
for the situation of Fig.~\protect\ref{fig:f03} (low coupling):
circuit A is shown in black and circuit C in red.
The corresponding temporal evolution for circuits
A (black) and B (red) is shown in plots (c) and (d). } \label{fig:f04}
\end{figure}

Figure~\ref{fig:f04} displays the time series of $V_1$ and $V_2$
for the three circuits, plotted in pairs for the sake of
comparison and keeping $R_{\rm coup}=20$~k$\Omega$. Figures~\ref{fig:f04}(a,b) show the outputs of circuits
A and B, which are clearly in antiphase. Furthermore, $V_1^A$ and
$V_1^B$ have different offsets, a fact that is reflected in the
phase space representation (Fig.~\ref{fig:f03}) in the form of
oscillations around two different
unstable fixed points. Figures~\ref{fig:f04}(c,d) show that circuits A and B remain
unsynchronized, a fact also observed for circuits B and C (not shown here).
Thus, the central circuit relays a state of antiphase
synchronization between the two outer ones, but remains
unsynchronized with them.

\begin{figure}[htb]
\centering
\includegraphics[width=60mm,clip]{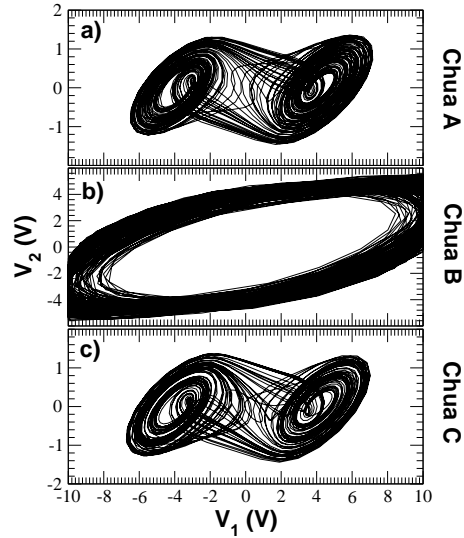}
\centering
\caption[2]{
Phase space dynamics of circuits A, B and C for low coupling, corresponding
to a coupling resistance $R_{\rm coup}= 20$~k$\Omega$. The conditions
are the same as in Fig.~\protect\ref{fig:f04}, but for a different
experimental realization (i.e. different initial conditions), which leads to a different dynamical state.}
\label{fig:f05}
\end{figure}

\begin{figure}[htb]
\centering
\includegraphics[width=80mm,clip]{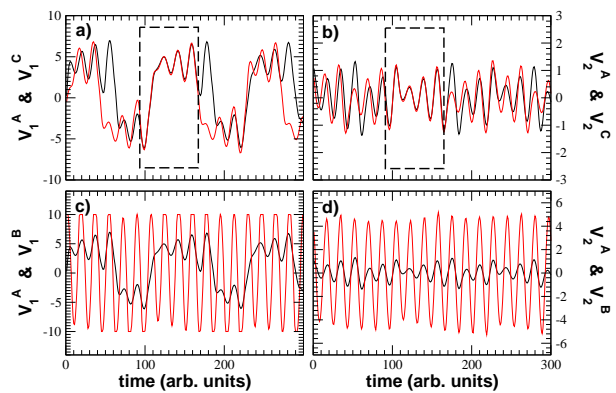}
\centering \caption[2]{
Temporal evolution of $V_1$ (a) and $V_2$ (b) of the outer circuits
for the situation of Fig.~\protect\ref{fig:f05} (low coupling):
circuit A is shown in black and circuit C in red.
The corresponding temporal evolution for circuits
A (black) and B (red) is shown in plots (c) and (d). 
The input voltage of the
data acquisition card is limited to $10V$. } \label{fig:f06}
\end{figure}

As we have mentioned above, the antiphase synchronized state described above
coexists with another possible behavior of the system. In different realizations of the
experiment (corresponding to different initial conditions),
the two outer circuits exhibit a double-scroll chaotic
attractor, while the central system oscillates in a quasi limit cycle 
(see the phase space representations in Fig.~\ref{fig:f05}).
We note that oscillations at the central circuit exceed the value
of $10$~V, which is the limit of the data acquisition card. This fact
that does not affect the dynamics of the system, but limits the
observable values of $V_1$.
Figure~\ref{fig:f06} shows the time series corresponding to
the dynamics of Fig.~\ref{fig:f05}, revealing synchronization episodes
between the outer circuits A and C [plots (a) and (b)] while the central one
performs almost periodic oscillations and remains unsynchronized with
the the outer circuits [(c) and (d)]. We ascribe the loss of
synchronization between the outer circuits to the interplay
between the intrinsic noise of the electronic circuits and the low
coupling. As in the previous state, the central Chua acts as a
relay between the outer systems in order to achieve
synchronization, but it does not participate in the synchronous
states between them.

\begin{figure}[htb]
\includegraphics[width=60mm,clip]{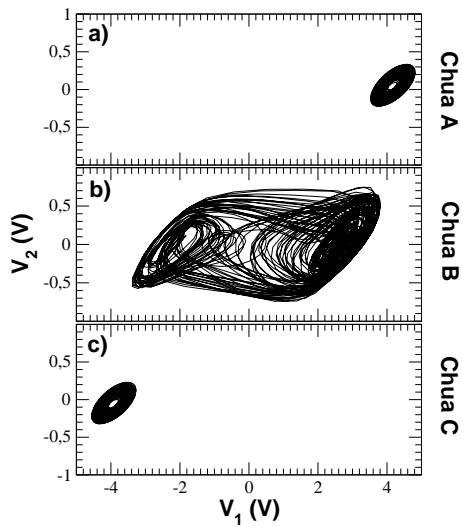}
\caption[2]{
Phase space dynamics of circuits A, B and C for intermediate coupling, corresponding
to a coupling resistance $R_{\rm coup}= 10$~k$\Omega$.}
\label{fig:f07}
\end{figure}

\begin{figure}[htb]
\includegraphics[width=80mm,clip]{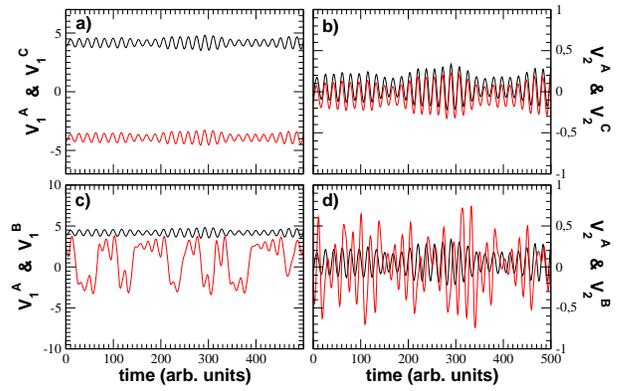}
\caption[2]{
Temporal evolution of $V_1$ (a) and $V_2$ (b) of the outer circuits
for the situation of Fig.~\protect\ref{fig:f07} (intermediate coupling):
circuit A is shown in black and circuit C in red.
The corresponding temporal evolution for circuits
A (black) and B (red) is shown in plots (c) and (d).} \label{fig:f08}
\end{figure}

We now increase coupling strength by reducing the coupling resistance. For
$R_{\rm coup}=10$~k$\Omega$, i.e. intermediate coupling, the collective
dynamics of the array changes qualitatively. The new situation is depicted in
Fig.~\ref{fig:f07}: the central circuit
develops a double-scroll chaotic attractor, while the two outer
ones exhibit a tight single-scroll attractor. The situation might look
similar to the low-coupling case shown in Figs.~\ref{fig:f03}-\ref{fig:f04},
since the two outer circuits oscillate around two
different unstable fixed points, while the
central system has a different trajectory in phase space.
Nevertheless differences arise when looking at the time series,
shown in Fig.~\ref{fig:f08}. In this case, circuits A and C remain
synchronized in phase [Figs.~\ref{fig:f08}(a,b)], although $V_1^A$ and
$V_1^C$ have different offsets. This type of synchronized behavior was
not observed in the case of low coupling, where the two outer
circuits operated in antiphase. Again, the central
circuit acts as an information relay, but remains unsynchronized with
the two outer circuits [Figs.~\ref{fig:f08}(c,d)].

\begin{figure}[htb]
\includegraphics[width=60mm,clip]{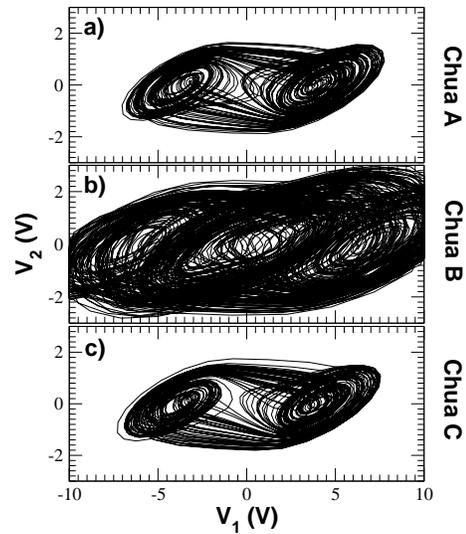}
\caption[2]{
Phase space dynamics of circuits A, B and C for high coupling, corresponding
to a coupling resistance $R_{\rm coup}= 2$~k$\Omega$.
}
\label{fig:f09}
\end{figure}

Finally, we increase the coupling even further. For
coupling resistances around $R_{\rm coup}\sim2$~k$\Omega$, which
correspond to reasonably high couplings, we observe a new change
in the dynamics of the array. Figure~\ref{fig:f09} shows the
phase space portraits of the three coupled circuits in this case. We can see
that circuits A and C exhibit a similar double-scroll chaotic
attractor. It is worth noting the shape of the chaotic
attractor of the central circuit, which is again different from that of the outer
ones. In this particular case, however, the central circuit exhibits a triple scroll
attractor, a kind of attractor not possible to obtain with an
isolated Chua's circuit. In this way, this kind of bidirectional
coupling can be used as a technique to generate different chaotic
attractors, as also observed in Ref.~\cite{zhong98}.
\begin{figure}[htb]
\includegraphics[width=80mm,clip]{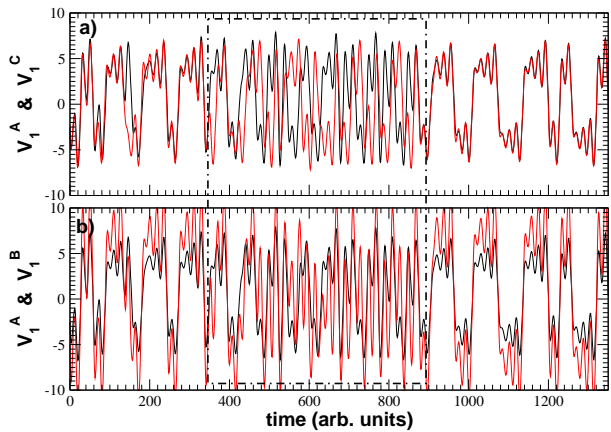}
\caption[2]{
Temporal evolution of $V_1$ (a) of the outer circuits (A and C)
for the situation of Fig.~\protect\ref{fig:f09} (high coupling):
circuit A is shown in black and circuit C in red.
The corresponding temporal evolution for circuits
A (black) and B (red) is shown in plot (b). Note the appearance
of unsynchronized windows (dotted box) between synchronization
regions. Identical synchronization $V_1^A=V_1^C$ is obtained  in (a), while
in (b) we observe generalized synchronization $V_1^B=aV_1^A$.
} \label{fig:f10}
\end{figure}
A closer look at the temporal evolution of $V_1$ and $V_2$ would
allow us to explain the appearance of the triple scroll attractor and to 
decide whether the dynamics of the systems are
synchronized or not. Figure~\ref{fig:f10}(a) shows that
circuits A and C exhibit identical synchronization, i.e.
$V_1^A=V_1^C$ (and $V_2^A=V_2^C$), coexisting with windows of unsynchronized behaviour. 
At the same time, synchronization
between circuit B (the one with the triple attractor) and the outer ones follows the expression $V_1^B=a
V_1^A$ (and $V_2^B=b V_2^A$), where $a$ (and $b$) is a constant [see synchronization region in Figure~\ref{fig:f10}(b)]. As
in the case of the outer circuits, synchronization coexists with states of unsynchronized behaviour and
it is during these transients when the system B lies within the central basin of the triple
scroll attractor. This kind of synchronization observed between the unsynchronized states, i.e. $V_1^B=a
V_1^A$, is known as generalized synchronization.
The same results are obtained when comparing circuits B and C (not
shown here). This is an experimental example of the coexistence of
generalized (between the central and outer circuits) and identical synchronization (between the outer circuits) 
in a chain of coupled chaotic oscillators.

\section{Conclusions}

In summary, we have investigated the dynamics of three bidirectionally
coupled Chua's circuits, analyzing the influence of the coupling strength
in the synchronization of these chaotic oscillators. For low coupling
strengths we observe coexistence of two states, which depend on the
initial conditions. One state corresponds to antiphase synchronized dynamics at
the outer systems, combined with chaotic unsynchronized dynamics with the
central one. In the other state we observe episodes of synchronization
between the outer systems while the central one has periodic oscillations,
and acts as a relay between the two outer systems. For intermediate
couplings, generalized synchronization between the outer circuits is obtained,
while the central system remains unsynchronized. Finally, for high
enough couplings, we observe coexistence of identical synchronization
between the outer circuits and generalized synchronization between
the central system and the outer ones. Furthermore, we report the
appearance of a triple scroll attractor at the central Chua's circuit, a chaotic
attractor not possible to obtain in isolated systems.

\section*{Acknowledgments}
We thank Ra\'{u}l Vicente and Ingo Fischer for fruitful
discussions. We acknowledge financial support from MEC (Spain) and
FEDER under Projects No. BFM2003-07850,
FIS2004-00953, TEC2005-007799, and from the Generalitat de Catalunya.

\appendix 

\section{The Chua's circuit} \label{app:circ}

\begin{figure}[htb]
\centering
\includegraphics[width=70mm,clip]{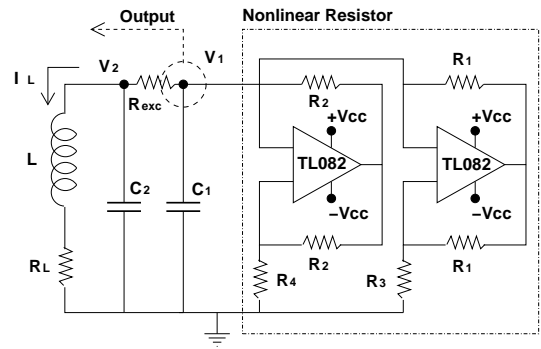}
\caption{
Description of the Chua circuit, which is built with two TL082
operational amplifiers and passive
electronic components of values: $V_{cc}=15$~V, $R_1=219.1$~$\Omega$, 
$R_2=21.92$~k$\Omega$,
$R_3=2.185$~k$\Omega$,  $R_4=3.28$~k$\Omega$, $R_L=23$~$\Omega$,
$C_1=9.8$~nF, $C_2=99.8$~nF, $L=19.54$~mH.
We set $R_{\rm exc}=1.73$~k$\Omega$ in order
to have chaotic dynamics. $V_1$ and/or $V_2$ correspond to the outputs of the circuit,
which are coupled to the other circuits through a voltage follower.
}
\label{fig:chua}
\end{figure}

Figure~\ref{fig:chua} shows a detailed description of the Chua's circuit used in
this work.
A nonlinear resistor is connected to a set of passive electronic components (R,L,C).
We have systematically studied the dynamical ranges of the circuit when
$R_{\rm exc}$ is modified, observing stable, periodic, excitable and chaotic dynamics.
Among all of them, we drive the circuit to have chaotic dynamics by
setting $R_{\rm exc}=1.73$~k$\Omega$. Under these conditions, the
dynamics of the circuit in the phase space given by ($V_1$,$V_2$) lies in a
single-scroll chaotic attractor (See Fig.~\ref{fig:f01}).
The output of the circuit ($V_1$ or $V_2$) is sent to the other circuits (with the
same characteristics) via
a voltage follower, in order to guarantee unidirectional injection (see
Appendix~\ref{app:coup} below for details on the coupling implementation).

The dynamics of the circuit are described by the equations \cite{kennedy92}:
\begin{eqnarray}
& &C_1\frac{dV_1}{dt} =
\frac{V_2-V_1}{R_exc}-g(V_1,V_{cc})
\\
& &C_2\frac{dV_2}{dt} =
\frac{V_1-V_2}{R_exc}+I_L
\\
& &L\frac{dI_L}{dt} =
-V_2-R_L I_L
\end{eqnarray}


\section{Coupling implementation} \label{app:coup}

\begin{figure}[htb]
\centering
\includegraphics[width=80mm,clip]{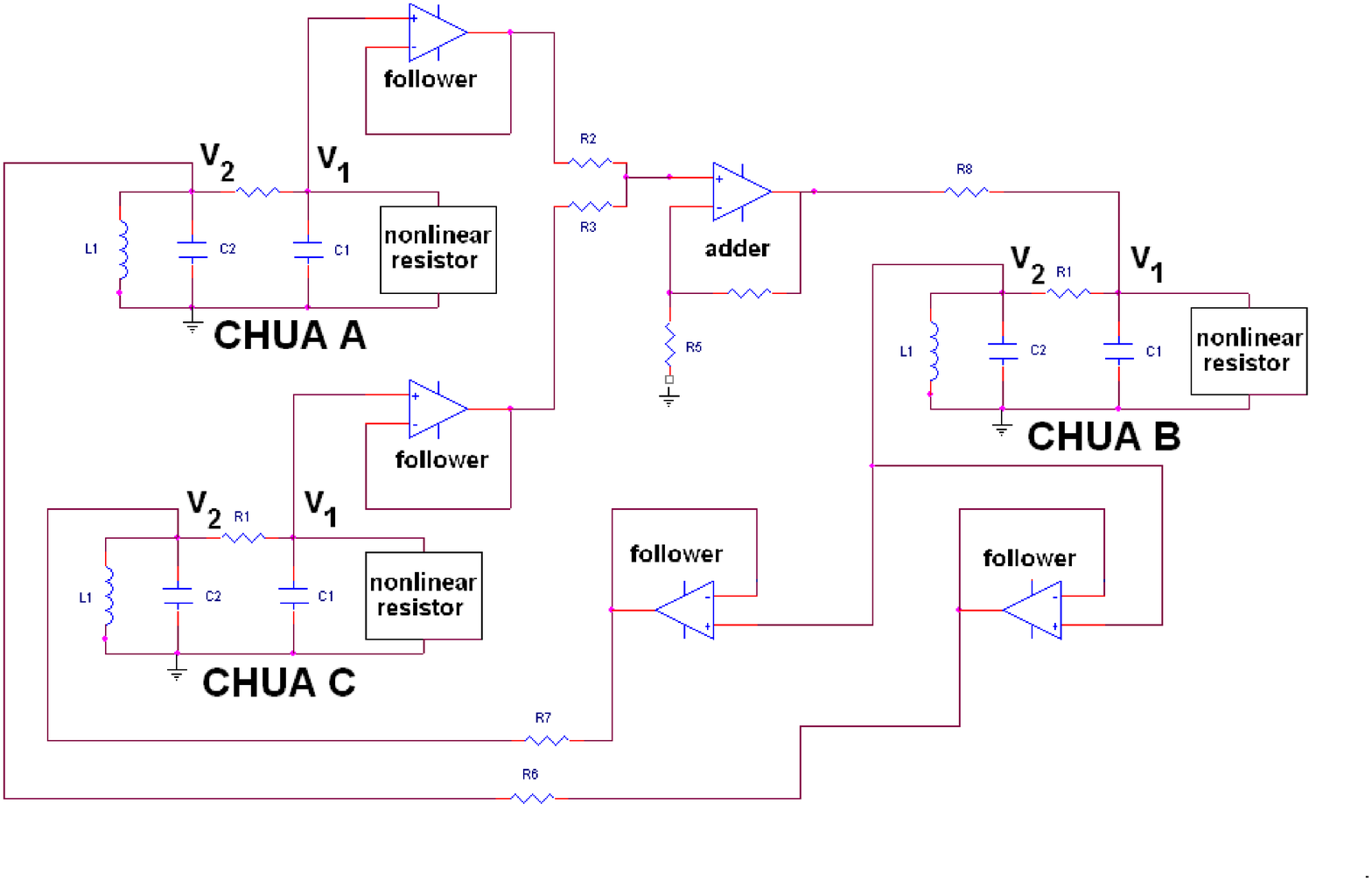}
\caption{
Description of the coupling connections between the three Chua's circuits.
TL082 operational amplifiers are used in the construction of the voltage
followers and adders.
}
\label{fig:fcoup}
\end{figure}

Figure~\ref{fig:fcoup} displays the architecture used to
bidirectionally couple the three Chua circuits (schematically described
in Fig.~\ref{fig:f01}). For the outer circuits, marked as A and C in
Fig.~\ref{fig:f01}, the output variable is $V_1$, which is sent to the central circuit (B)
through a voltage adder. The coupling strength is controlled by
adjusting the value of $R_8$. At the same time, $V_2$ coming from circuit B is
the input variable, which comes trough two voltage followers (one for
each outer circuit) and whose strength is adjusted by $R_7$ and $R_6$. It
is worth noting that $R_8=R_7=R_6$ throughout the experiment, in order
to keep the same coupling strength between all pairs of circuits.

\end{document}